\documentclass[]{spie}  

\usepackage{amsmath,amsfonts,amssymb}
\usepackage{graphicx}
\usepackage[colorlinks=true, allcolors=blue]{hyperref}
\usepackage{glossaries}
\usepackage{soul}
\usepackage{subfigure}

\usepackage{tabularx}
\newcolumntype{L}[1]{>{\raggedright\let\newline\\\arraybackslash\hspace{0pt}}m{#1}} 

\makeatletter
\let\jnl@style=\rm
\def\ref@jnl#1{{\jnl@style#1}}
\def\aj{\ref@jnl{AJ}}                   
\def\actaa{\ref@jnl{Acta Astron.}}      
\def\araa{\ref@jnl{ARA\&A}}             
\def\apj{\ref@jnl{ApJ}}                 
\def\apjl{\ref@jnl{ApJ}}                
\def\apjs{\ref@jnl{ApJS}}               
\def\ao{\ref@jnl{Appl.~Opt.}}           
\def\apss{\ref@jnl{Ap\&SS}}             
\def\aap{\ref@jnl{A\&A}}                
\def\aapr{\ref@jnl{A\&A~Rev.}}          
\def\aaps{\ref@jnl{A\&AS}}              
\def\azh{\ref@jnl{AZh}}                 
\def\baas{\ref@jnl{BAAS}}               
\def\bac{\ref@jnl{Bull. astr. Inst. Czechosl.}}
\def\caa{\ref@jnl{Chinese Astron. Astrophys.}}
\def\cjaa{\ref@jnl{Chinese J. Astron. Astrophys.}}
\def\icarus{\ref@jnl{Icarus}}           
\def\jcap{\ref@jnl{J. Cosmology Astropart. Phys.}}
\def\jrasc{\ref@jnl{JRASC}}             
\def\memras{\ref@jnl{MmRAS}}            
\def\mnras{\ref@jnl{MNRAS}}             
\def\na{\ref@jnl{New A}}                
\def\nar{\ref@jnl{New A Rev.}}          
\def\pra{\ref@jnl{Phys.~Rev.~A}}        
\def\prb{\ref@jnl{Phys.~Rev.~B}}        
\def\prc{\ref@jnl{Phys.~Rev.~C}}        
\def\prd{\ref@jnl{Phys.~Rev.~D}}        
\def\pre{\ref@jnl{Phys.~Rev.~E}}        
\def\prl{\ref@jnl{Phys.~Rev.~Lett.}}    
\def\pasa{\ref@jnl{PASA}}               
\def\pasp{\ref@jnl{PASP}}               
\def\pasj{\ref@jnl{PASJ}}               
\def\rmxaa{\ref@jnl{Rev. Mexicana Astron. Astrofis.}}%
\def\qjras{\ref@jnl{QJRAS}}             
\def\skytel{\ref@jnl{S\&T}}             
\def\solphys{\ref@jnl{Sol.~Phys.}}      
\def\sovast{\ref@jnl{Soviet~Ast.}}      
\def\ssr{\ref@jnl{Space~Sci.~Rev.}}     
\def\zap{\ref@jnl{ZAp}}                 
\def\nat{\ref@jnl{Nature}}              
\def\iaucirc{\ref@jnl{IAU~Circ.}}       
\def\aplett{\ref@jnl{Astrophys.~Lett.}} 
\def\apspr{\ref@jnl{Astrophys.~Space~Phys.~Res.}}
\def\bain{\ref@jnl{Bull.~Astron.~Inst.~Netherlands}} 
\def\fcp{\ref@jnl{Fund.~Cosmic~Phys.}}  
\def\gca{\ref@jnl{Geochim.~Cosmochim.~Acta}}   
\def\grl{\ref@jnl{Geophys.~Res.~Lett.}} 
\def\jcp{\ref@jnl{J.~Chem.~Phys.}}      
\def\jgr{\ref@jnl{J.~Geophys.~Res.}}    
\def\jqsrt{\ref@jnl{J.~Quant.~Spec.~Radiat.~Transf.}}
\def\memsai{\ref@jnl{Mem.~Soc.~Astron.~Italiana}}
\def\nphysa{\ref@jnl{Nucl.~Phys.~A}}   
\def\physrep{\ref@jnl{Phys.~Rep.}}   
\def\physscr{\ref@jnl{Phys.~Scr}}   
\def\planss{\ref@jnl{Planet.~Space~Sci.}}   
\def\procspie{\ref@jnl{Proc.~SPIE}}   

\let\amp=\&
\makeatother

\title{A review of simulation and performance modeling tools for the Roman coronagraph instrument}
\author[a]{Ewan S. Douglas} 
\author[a]{Jaren N. Ashcraft}
\author[b]{Ruslan Belikov}
\author[c]{John Debes}
\author[d,f]{Jeremy Kasdin}
\author[e]{John Krist}
\author[f]{Brianna I Lacy}
\author[h]{Bijan Nemati}
\author[a]{Kian Milani}
\author[g]{Leonid Pogorelyuk}
\author[e]{A J Eldorado Riggs}
\author[i]{Dmitry Savransky}
\author[b]{Dan Sirbu}
\affil[a]{University of Arizona,  Steward Observatory, Tucson, AZ, USA}
\affil[b]{NASA Ames Research Center, Moffett Field, CA, USA}
\affil[c]{Space Telescope Science Institute, Baltimore, MD, USA}
\affil[d]{University of California San Francisco, San Francisco, CA, USA}
\affil[e]{Jet Propulsion Laboratory, California Institute of Technology, Pasadena, CA, USA}
\affil[f]{Princeton University, Department of Astrophysical Sciences, Princeton, NJ, USA}
\affil[g]{Massachusetts Institute of Technology, Cambridge, MA, USA}
\affil[h]{University of Alabama in Huntsville, Huntsville, AL, USA}
\affil[i]{Cornell University, Ithaca, NY, USA}

\authorinfo{Further author information: (Send correspondence to E.S.D.)\\E.S.D,: E-mail: douglase@arizona.edu}

\begin{document} 
\maketitle


\newacronym{AU}{AU}{astronomical Unit [1.5e11 m]}  
\newacronym{pc}{pc}{parsec}
\newacronym{mas}{mas}{milliarcsecond}
\newacronym{nm}{nm}{nanometer}
\newacronym{CTE}{CTE}{coefficient of thermal expansion}
\newacronym{sqarc}{$as^2$}{square arcsecond}

\newacronym{smc}{SMC}{Small Magellanic Cloud}
\newacronym{lmc}{LMC}{Large Magellanic Cloud}
\newacronym{ism}{ISM}{interstellar medium}
\newacronym{mw}{MW}{Milky Way}
\newacronym{epseri}{$\epsilon$ Eri}{Epsilon Eridani}
\newacronym{EKB}{EKB}{Edgeworth-Kuiper Belt}

\newacronym{CFR}{CFR}{Complete Frequency Redistribution}

\newacronym{nasa}{NASA}{National Aeronautics and Space Agency}
\newacronym{esa}{ESA}{European Space Agency}
\newacronym{omi}{OMI}{\textit{Optical Mechanics Inc.}}
\newacronym{gsfc}{GSFC}{\gls{nasa} Goddard Space Flight Center}
\newacronym{stsci}{STScI}{Space Telescope Science Institute}
\newacronym{nsroc}{NSROC}{\gls{nasa} Sounding Rocket Operations Contract}
\newacronym{wff}{WFF}{\gls{nasa} Wallops Flight Facility}
\newacronym{wsmr}{WSMR}{White Sands Missile Range}

\newacronym{irac}{IRAC}{Infrared Array Camera}
\newacronym[plural=CCDs, firstplural=charge-coupled devices (CCDs)]{ccd}{CCD}{charge-coupled device}
\newacronym[plural=EMCCDs, firstplural=electron multiplying charge-coupled devices (EMCCDs)]{EMCCD}{EMCCD}{electron multiplying charge-coupled device}

\newacronym{DM}{DM}{Deformable Mirror}
\newacronym{MCP}{MCP}{ Microchannel Plate }
\newacronym{ipc}{IPC}{Image Proportional Counter}
\newacronym{cots}{COTS}{Commercial Off-The-Shelf}
\newacronym{ISR}{ISR}{incoherent scatter radar}
\newacronym{atcamera}{AT}{angle tracker}
\newacronym{MEMS}{MEMS}{microelectromechanical systems}
\newacronym{QE}{QE}{quantum efficiency}
\newacronym{RTD}{RTD}{Resistance Temperature Detector}
\newacronym{PID}{PID}{Proportional-Integral-Derivative}
\newacronym{PRNU}{PRNU}{photo response non-uniformity}
\newacronym{DSNU}{PRNU}{dark signal non-uniformity}
\newacronym{CMOS}{CMOS}{complementary metal–oxide–semiconductor}
\newacronym{TRL}{TRL}{technology readiness level}
\newacronym{swap}{SWaP}{Size, Weight, and Power}
\newacronym{ConOps}{ConOps}{concept of operations}
\newacronym{NRE}{NRE}{non-recurring engineering}
\newacronym{CBE}{CBE}{current best estimate}

\newacronym{FOV}{FOV}{field-of-view}
\newacronym{NIR}{NIR}{near-infrared}
\newacronym{PV}{PV}{Peak-to-Valley}
\newacronym{MRF}{MRF}{Magnetorheological finishing}
\newacronym{AO}{AO}{Adaptive Optics}
\newacronym{TTP}{TTP}{tip, tilt, and piston}
\newacronym{FPS}{FPS}{fine pointing system}
\newacronym{SHWFS}{SHWFS}{Shack-Hartmann Wavefront Sensor}
\newacronym{OAP}{OAP}{off-axis parabola}
\newacronym{LGS}{LGS}{laser guide star}
\newacronym{WFCS}{WFCS}{wavefront control system}
\newacronym{OPD}{OPD}{optical path difference}
\newacronym{UA}{UA}{University of Arizona}
\newacronym{MEL}{MEL}{Master Equipment List}
\newacronym{GEO}{GEO}{low-earth orbit}
\newacronym{LEO}{LEO}{geosynchronous orbit}
\newacronym{EFC}{EFC}{electric-field conjugation}
\newacronym{LDFC}{LDFC}{linear dark field control}
\newacronym{DAC}{DAC}{digital-to-analog converter}
\newacronym{FEA}{FEA}{finite element analysis}

\newacronym{SiC}{SiC}{Silicon Carbide}
\newacronym{ESPA}{ESPA}{EELV Secondary Payload Adapter}
\newacronym{EEID}{EEID}{Earth-equivalent Insolation Distance, the distance from the star where the incident energy density is that of the Earth received from the Sun}
\newacronym{LLOWFS}{LLOWFS}{Lyot low-order wavefront sensor}
\newacronym{STOP}{STOP}{Structural-Thermal-Optical-Performance}

\newacronym{resel}{resel}{resolution element}

\newacronym{acs}{ACS}{Attitude Control System}
\newacronym{orsa}{ORSA}{Ogive Recovery System Assembly}
\newacronym{gse}{GSE}{Ground Station Equipment}
\newacronym{FSM}{FSM}{Fast Steering Mirror}

\newacronym{WFS}{WFS}{wavefront sensor}
\newacronym{LSI}{LSI}{Lateral Shearing Interferometer}
\newacronym{VVC}{VVC}{Vector Vortex Coronagraph}
\newacronym{VNC}{VNC}{Visible Nulling Coronagraph}
\newacronym{CGI}{CGI}{Coronagraph Instrument}
\newacronym{IWA}{IWA}{Inner Working Angle}
\newacronym{OWA}{OWA}{Outer Working Angle}
\newacronym{NPZT}{N-PZT}{Nuller piezoelectric transducer}
\newacronym{ZWFS}{ZWFS}{Zernike wavefront sensor}
\newacronym{SPC}{SPC}{Shaped Pupil Coronagraph}
\newacronym{HLC}{HLC}{Hybrid-Lyot Coronagraph}
\newacronym{ADI}{ADI}{angular differential imaging}
\newacronym{RDI}{RDI}{reference differential imaging}
\newacronym{LOWFSC}{LOWFS/C}{low-order wavefront sensing and control}
\newacronym{HOWFSC}{HOWFS/C}{high-order wavefront sensing and control}
\newacronym{WFSC}{WFSC}{wavefront sensing and control}

\newacronym{HST}{HST}{Hubble Space Telescope}
\newacronym{GPS}{GPS}{Global Positioning System}
\newacronym{ISS}{ISS}{International Space Station}
\newacronym[description=Advanced CCD Imaging Spectrometer]{acis}{ACIS}{Advanced \gls{ccd} Imaging Spectrometer}
\newacronym{stis}{STIS}{\textit{Space Telescope Imaging Spectrograph}}
\newacronym{mcp}{MCP}{Microchannel Plate}
\newacronym{jwst}{JWST}{$\textit{James Webb Space Telescope}$}
\newacronym{fuse}{FUSE}{$\textit{FUSE}$}
\newacronym{galex}{GALEX}{$\textit{Galaxy Evolution Explorer}$}
\newacronym{spitzer}{Spitzer}{$\textit{Spitzer Space Telescope}$}
\newacronym{mips}{MIPS}{Multiband Imaging Photometer for \gls{spitzer}}
\newacronym{gissmo}{GISSMO}{Gas Ionization Solar Spectral Monitor}
\newacronym{iue}{IUE}{International Ultraviolet Explorer}
\newacronym{spinr}{SPINR}{$\textit{Spectrograph for Photometric Imaging with Numeric Reconstruction}$}
\newacronym{imager}{IMAGER}{$\textit{Interstellar Medium Absorption Gradient Experiment Rocket}$}
\newacronym{TPF-C}{TPF-C}{Terrestrial Planet Finder Coronagraph}
\newacronym{RAIDS}{RAIDS}{Atmospheric and Ionospheric Detection System }
\newacronym{mama}{MAMA}{Multi-Anode Microchannel Array}
\newacronym{ATLAST}{ATLAST}{Advanced Technology Large Aperture Space Telescope}
\newacronym{PICTURE}{PICTURE}{Planet Imaging Concept Testbed Using a Rocket Experiment}
\newacronym{LITES}{LITES}{Limb-imaging Ionospheric and Thermospheric
Extreme-ultraviolet Spectrograph}
\newacronym{LBT}{LBT}{Large Binocular Telescope}
\newacronym{LBTI}{LBTI}{Large Binocular Telescope Interferometer}
\newacronym{KIN}{KIN}{Keck Interferometer Nuller}
\newacronym{SHARPI}{SHARPI}{Solar High-Angular Resolution Photometric Imager}
\newacronym{IRAS}{IRAS}{Infrared Astronomical Satellite}
\newacronym{HARPS}{HARPS}{High Accuracy Radial velocity Planetary}
\newacronym{hstSTIS}{STIS}{Space Telescope Imaging Spectrograph}
\newacronym{spitzerIRAC}{IRAC}{Infrared Array Camera}
\newacronym{spitzerMIPS}{MIPS}{Multiband Imaging Photometer for Spitzer}
\newacronym{spitzerIRS}{IRS}{Infrared Spectrograph}
\newacronym{CHARA}{CHARA}{Center for High Angular Resolution Astronomy}
\newacronym{wfirst-afta}{WFIRST-AFTA}{Wide-Field InfrarRed Survey
Telescope-Astrophysics Focused Telescope Assets}
\newacronym{GPI}{GPI}{Gemini Planet Imager}
\newacronym{WFIRST}{Roman}{Nancy Grace Roman Space Telescope}
\newacronym{HabEx}{HabEx}{Habitable Exoplanet Observatory Mission Concept}
\newacronym{LUVOIR}{LUVOIR}{Large UV/Optical/Infrared Surveyor}
\newacronym{FGS}{FGS}{Fine Guidance Sensor}
\newacronym{STIS}{STIS}{Space Telescope Imaging Spectrograph}
\newacronym{MGHPCC}{MGHPCC}{Massachusetts Green High Performance
Computing Center}
\newacronym{WISE}{WISE}{Wide-field Infrared Survey Explorer}
\newacronym{ALMA}{ALMA}{Atacama Large Millimeter Array}
\newacronym{GRAIL}{GRAIL}{Gravity Recovery and Interior Laboratory}
\newacronym{jwstNIRCam}{NIRCam}{near-\gls{IR}-camera}
\newacronym{jwstMIRI}{MIRI}{Mid-Infrared Instrument}

\newacronym{AURIC}{AURIC}{The Atmospheric Ultraviolet Radiance Integrated Code} 
\newacronym{FFT}{FFT}{Fast Fourier Transform  }
\newacronym{MODTRAN}{MODTRAN   }{ MODerate resolution atmospheric TRANsmission }
\newacronym{idl}{IDL}{$\textit {Interactive Data Language}$}
\newacronym[sort=NED,description=NASA/IPAC Extragalactic Database]{ned}{NED}{\gls{nasa}/\gls{ipac} Extragalactic Database}
\newacronym{iraf}{IRAF}{Image Reduction and Analysis Facility}
\newacronym{wcs}{WCS}{World Coordinate System}
\newacronym{pegase}{PEGASE}{$\textit{Projet d'Etude des GAlaxies par Synthese Evolutive}$}
\newacronym{dirty}{DIRTY}{$\textit{DustI Radiative Transfer, Yeah!}$}
\newacronym{CUDA}{CUDA}{Compute Unified Device Architecture}
\newacronym{KLIP}{KLIP}{Karhunen-Lo`eve Image Processing}

\newacronym{MSIS}{MSIS}{Mass Spectrometer Incoherent Scatter Radar}
\newacronym{nmf2}{$N_m$}{F2-Region Peak density}
\newacronym{hmf2}{$h_m$}{F2-Region Peak height}
\newacronym{H}{$H$}{F2-Region Scale Height}

\newacronym{isr}{ISR}{Incoherent Scatter Radar}
\newacronym[description=TLA Within Another Acronym]{twaa}{TWAA}{\gls{tla} Within Another Acronym}
\newacronym[plural=SNe, firstplural=Supernovae (SNe)]{sn}{SN}{Supernova}
\newacronym{EUV}{EUV}{Extreme-Ultraviolet }
\newacronym{EUVS}{EUVS}{\gls{EUV} Spectrograph}
\newacronym{F2}{F2}{Ionospheric Chapman F Layer }
\newacronym{F10.7}{F10.7}{ 10.7 cm radio flux [10$^{-22}$ W m$^{-2}$ Hz$^{-1}$]  }
\newacronym{FUV}{FUV}{far-ultraviolet }
\newacronym{IR}{IR}{infrared}
\newacronym{MUV}{MUV}{mid-ultraviolet }
\newacronym{NUV}{NUV}{near-ultraviolet }
\newacronym{O$^+$}{O$^+$}{Singly Ionized Oxygen  Atom }
\newacronym{OI}{OI}{Neutral Atomic Oxygen Spectroscopic State }
\newacronym{OII}{OII}{Singly Ionized Atomic Oxygen Spectroscopic State }
\newacronym{PSF}{PSF}{point spread function}
\newacronym{$R_E$}{$R_E$}{Earth radii [$\approx$ 6400 km]  }
\newacronym{RV}{RV}{radial velocity}
\newacronym{UV}{UV}{ultraviolet }
\newacronym{WFE}{WFE}{wavefront error}
\newacronym{sed}{SED}{spectral energy distribution}
\newacronym{nir}{NIR}{near-infrared}
\newacronym{mir}{MIR}{mid-infrared}
\newacronym{ir}{IR}{infrared}
\newacronym{uv}{UV}{ultraviolet}
\newacronym[plural=PAHs, firstplural=Polycyclic Aromatic Hydrocarbons (PAHs)]{pah}{PAH}{Polycyclic Aromatic Hydrocarbon}
\newacronym{obsid}{OBSID}{Observation Identification}
\newacronym{SZA}{SZA}{Solar Zenith Angle}
\newacronym{TLE}{TLE}{Two Line Element set}
\newacronym{DOF}{DOF}{degrees-of-freedom}
\newacronym{PZT}{PZT}{lead zirconate titanate}
\newacronym{ADCS}{ADCS}{attitude determination and control system}
\newacronym{COTS}{COTS}{commercial off-the-shelf}
\newacronym{CDH}{C$\&$DH}{command and data handling}
\newacronym{EPS}{EPS}{electrical power system}

\newacronym{PCA}{PCA}{principal component analysis}
\newacronym{fwhm}{FWHM}{full-width-half maximum}
\newacronym{RMS}{RMS}{root mean squared}
\newacronym{RMSE}{RMSE}{root mean squared error}
\newacronym{MCMC}{MCMC}{Marcov chain Monte Carlo}
\newacronym{DIT}{DIT}{discrete inverse theory}
\newacronym{SNR}{SNR}{signal-to-noise ratio}
\newacronym{PSD}{PSD}{power spectral density}
\newacronym{NMF}{NMF}{non-negative matrix factorization}

\begin{abstract}

The Nancy Grace Roman Space Telescope Coronagraph Instrument (CGI) will be capable of characterizing exoplanets in reflected light and will demonstrate space technologies essential for future missions to take spectra of Earthlike exoplanets. As the mission and instrument move into the final stages of design, simulation tools spanning from depth of search calculators to detailed diffraction models have been created by a variety of teams.  We summarize these efforts, with a particular focus on publicly available datasets and software tools. These include speckle and point-spread-function models, signal-to-noise calculators, and science product simulations (e.g. predicted observations of debris disks and exoplanet spectra). This review is intended to serve as a reference to facilitate engagement with the technical and science capabilities of the CGI instrument. 

\end{abstract}


\section{INTRODUCTION}
\label{sec:intro}  
The \gls{WFIRST}\cite{spergel_wide-field_2015} \gls{CGI} is a technology demonstration\cite{noecker_coronagraph_2016,kasdin_wfirst_2018,douglas_wfirst_2018,bailey_wfirst_2019-1} that will use high-contrast imaging and spectroscopy (coronagraphy),  wavefront sensing, and wavefront control\cite{shi_low_2016,sidick_wfirst_2018}, to image planets in reflected light \cite{kasdin_wfirst_2018}.
Formerly known as the Wide-Field Infrared Survey Telescope, \gls{WFIRST} is orders of magnitude more sensitive than \gls{HST} or ground-based observatories\cite{bailey_wfirst_2019}.
Two primary coronagraph technologies will be tested, a \gls{HLC} coronagraph\cite{trauger_hybrid_2016} for high contrast small-\gls{FOV} ($<$0.5 as) imaging, a bow-tie \gls{SPC}\cite{balasubramanian_wfirst-afta_2015} for spectroscopy and a wide-\gls{FOV} \gls{SPC}.
The design and  preparation for \gls{CGI} has spawned many new modeling tools to accurately predict performance. 
This work attempts to summarizes the most common of these tools, both to serve as a roadmap for future potential users of \gls{CGI} and to aid other missions which may benefit from reuse of the many open source tools which have been shared by \gls{CGI} science and engineering teams. 
This review cannot completely capture the simulation work that has gone into developing Roman CGI but it is intended to serve as a starting reference for the community to engage with the technical and science capabilities of the instrument. 
Additionally, to better understand the context of these tools, we suggest previous works, as well as several papers in these proceedings, that provide detailed descriptions of the design reference mission and typical observing modes\cite{kasdin_wfirst_2018,bailey_wfirst_2019,poberezhskiy_cgi_2020,kasdin_wfirst_2020}.

\begin{table}[ht]
\renewcommand*{\arraystretch}{1.5}
    \centering
    \caption{Partial Listing of Open Source CGI Software Packages and Libraries}
    \label{tab:my_label}
    \footnotesize
    \begin{tabular}{L{3.75cm}|L{3.75cm}|c|p{6.25cm}}
    \hline
    Package & Application & References & URL\\
    \hline
    Observing Scenarios    & Complete Observation Simulation  & \citenum{krist_numerical_2015}& \url{https://roman.ipac.caltech.edu/sims/Coronagraph_public_images.html}\\
    PROPER & Diffraction Simulation & \citenum{krist_proper:_2007,krist_wfirst_2017} & \url{https://github.com/ajeldorado/proper-models/tree/master/wfirst_cgi}\\
     \verb+FALCO+  & Coronagraph Simulation   & \citenum{riggs2018falco1} & \url{https://github.com/ajeldorado/falco-matlab},  \url{https://github.com/ajeldorado/falco-python} \\
     Lightweight Coronagraph Simulator & Coronagraph Simulation & \citenum{pogorelyuk_effects_2020}& \url{https://github.com/leonidprinceton/LightweightSpaceCoronagraphSimulator}\\
     CZT-based Optical Propagation & Diffraction Simulation && \url{https://github.com/ARCExoplanetTechnologies/ACED}\\
           \verb+WebbPSF+ & Diffraction Simulation & \citenum{perrin_simulating_2012,perrin_updated_2014}  & \url{https://github.com/spacetelescope/webbpsf}\\

     MSWC & Binary Star Simulation & \citenum{dsirbu2017mswc,dsirbu2018RomanMSWC} & \url{https://github.com/ARCExoplanetTechnologies/MSWC}\\
     \verb+EXOSIMS+ & Mission Simulation & \citenum{savransky_exosims_2017,savransky_wfirst-afta_2016} &\url{https://github.com/dsavransky/EXOSIMS}\\
     Imaging Mission Database&Mission Planning&\citenum{savransky_exploration_2019} & \url{https://plandb.sioslab.com}\\
     Coronagraph convolved Debris Disks & Disk Simulation & \citenum{mennesson_wfirst_2018}& \url{https://roman.ipac.caltech.edu/sims/Circumstellar_Disk_Sims.html} \\
     Known debris disk simulated scenes & Disk Simulation &\citenum{chen_wfirst_nodate}&  \url{https://roman.ipac.caltech.edu/sims/Chen_WPS.html}\\
     direct-imaging-sims + model spectra & Target simulation &\citenum{lacy_characterization_2019, lacy_prospects_2020}  & \url{https://github.com/blacy/direct-imaging-sims},
    \url{https://www.astro.princeton.edu/~burrows/wfirst/index.html}\\
         Giant planet albedo spectra & Target simulation & \citenum{cahoy_exoplanet_2010}  &  \url{https://wfirst.ipac.caltech.edu/sims/Exoplanet_Characterization.html}\\
     \hline
    \end{tabular}
\end{table}

\section{Coronagraph Simulators}
\label{sec:packages}  
\subsection{Observing Scenarios}
The most detailed and physically realistic simulations of \gls{CGI} observations released to date have taken the form of numbered observing scenarios. 
These are referred to as OS$n$, e.g., OS9 for the ninth scenario.
Physical optics simulations of the instrument including optical surfaces, coronagraphs, and wavefront sensing and control are generated using PROPER\cite{krist2007proper,krist_wfirst_2018}.
The scenarios include speckle time series, derived from wavefront maps produced using \gls{STOP} modeling of the Roman observatory\cite{saini_impipeline_2017}.
The OS model outputs include noisy and noiseless datacubes of coronagraphic intensity images versus time, with injected planets and realistic observing scenarios that include reference stars, and off-axis \gls{PSF}s for injecting additional targets.
These scenario files are available to the public from IPAC.

\begin{figure}[ht]
    \centering
    \includegraphics{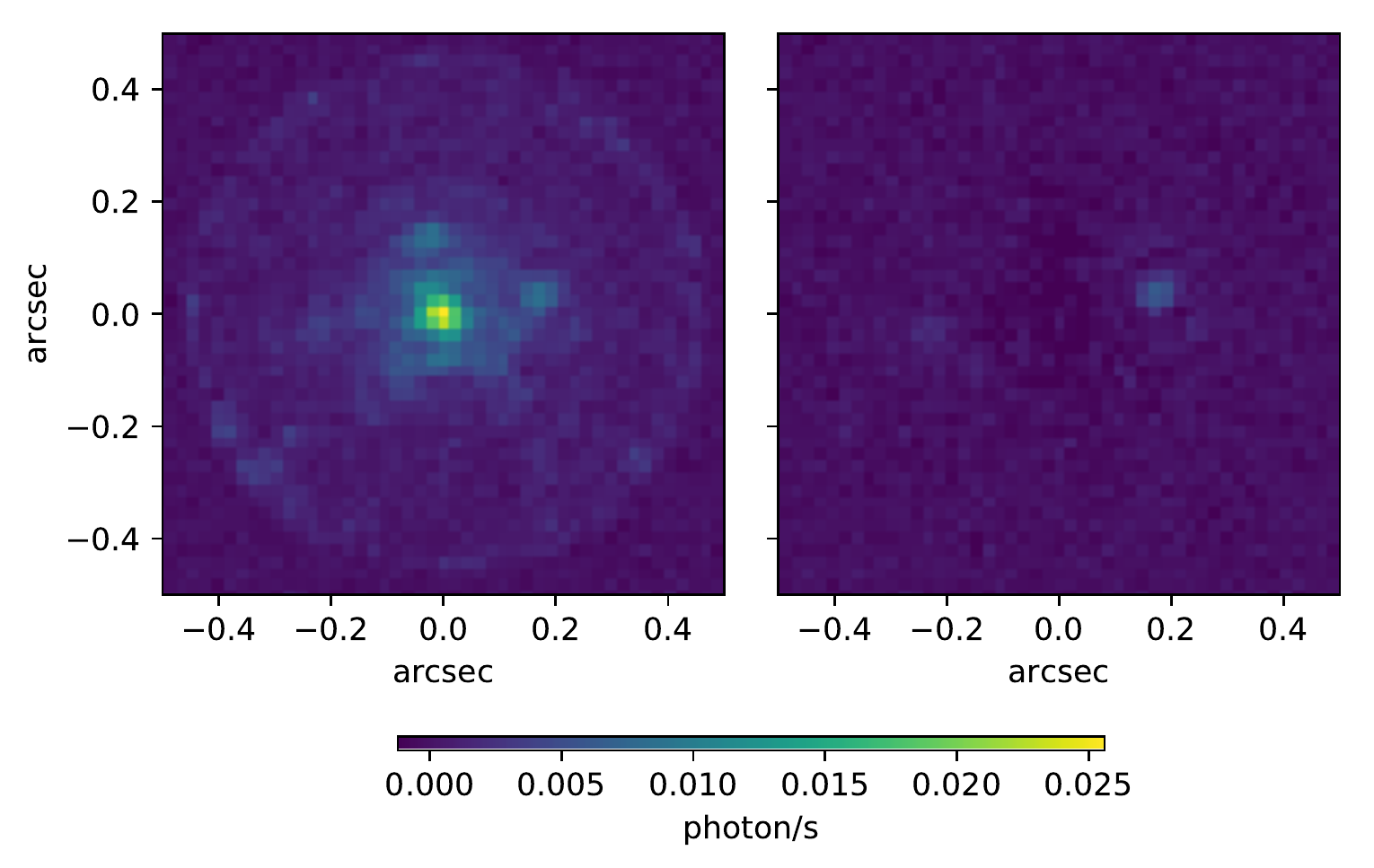}
    \caption{Images of 47 Uma c generated from publicly available OS9 simulations of the \gls{HLC}  including time dependent speckle and detector effects. Left: simulated image with injected planets. Right: the same scene after reference subtraction, showing injected planets at a much higher \gls{SNR}.}
    \label{fig:coronagraphs}
\end{figure}

\subsection{FALCO}
\begin{figure}[ht]
\centering
\subfigure[] { 
    \label{fig:before_wfsc}
    \includegraphics[width = .4\columnwidth,trim=.05in 0in .05in 0in]{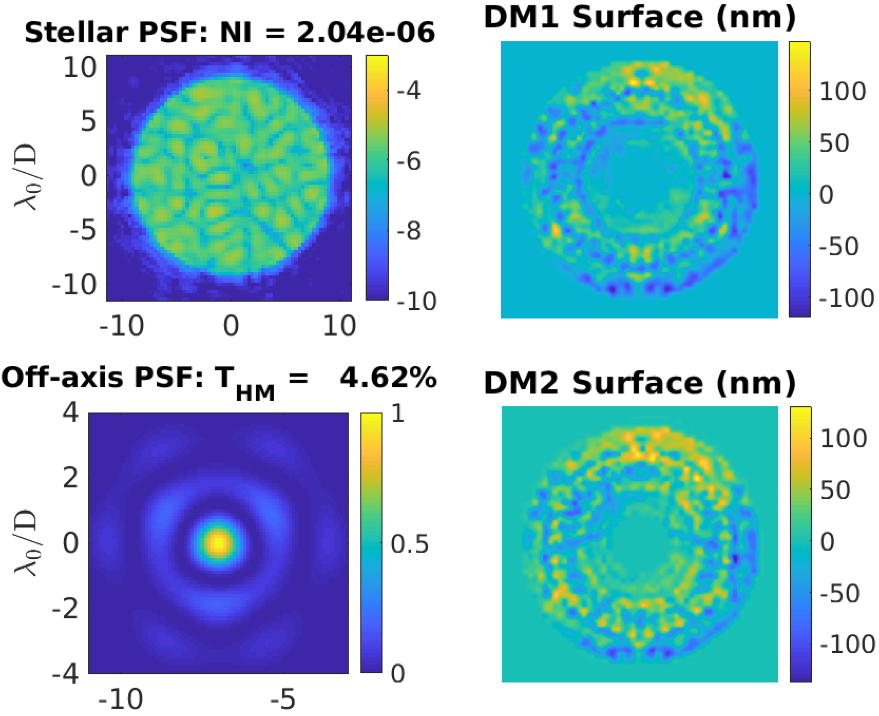}}
 \subfigure[] {
    \label{fig:after_wfsc}
    \includegraphics[width = .4\columnwidth,trim=.05in 0in .05in 0in]{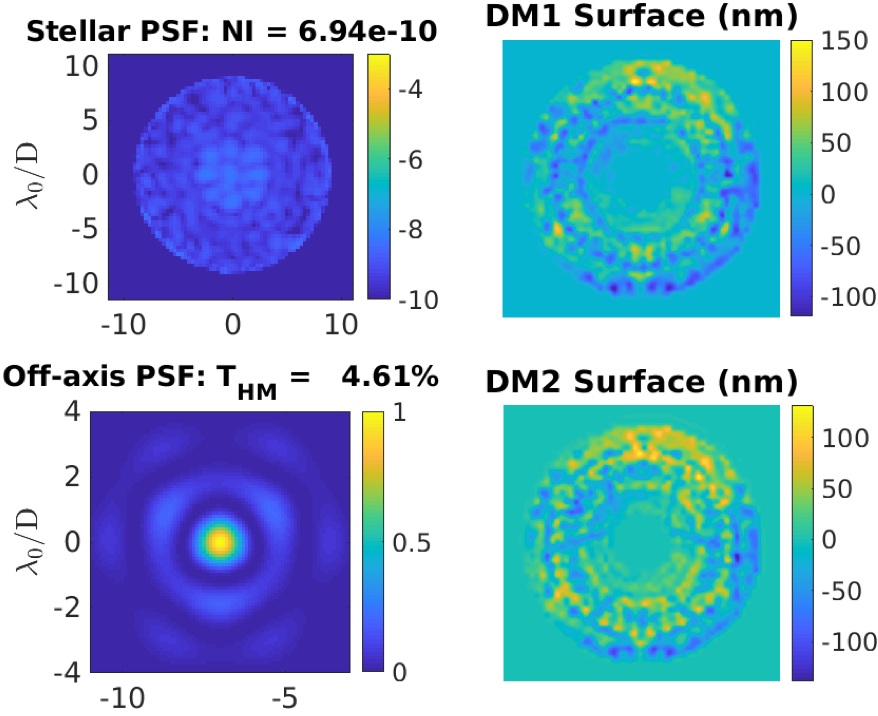}}   
\caption{Status report windows from FALCO before (a) and after (b) performing wavefront control using the CGI PROPER model as the truth model. The large starting shapes on the deformable mirrors are the HLC design settings combined with the settings for flattening the starting wavefront errors.}
\label{fig:FALCO}
\end{figure}

The FALCO\cite{riggs2018falco1} library provides a framework for running wavefront sensing and control algorithms in MATLAB and Python 3. FALCO can use a PROPER\cite{krist2007proper} model as its truth model in simulations, and examples using the CGI PROPER model are included in the publicly available FALCO repository. Example status window updates for the hybrid Lyot coronagraph are shown in Fig.~\ref{fig:FALCO}.

\subsection{Lightweight Space Coronagraph Simulator}
\begin{figure}[ht]
    \centering
    \includegraphics[width=0.9\textwidth]{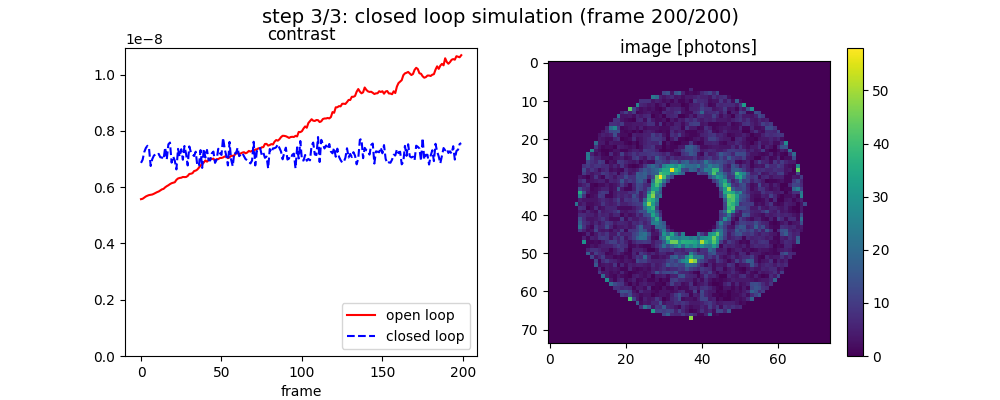}
    \caption{An example of contrast evolution in presence of various wavefront instabilities in a linearized model of Roman-CGI. Left: Closing the loop allows maintaining the contrast throughout the observation. The drift rate is exaggerated to illustrate dark hole maintenance in the worst-case scenario of random walk of Zernike coefficients. Right: A broadband photon-counts image of a single exposure (used to close the contrast loop). See Pogorelyuk et al.\cite{pogorelyuk_effects_2020} for details.}
    \label{fig:leonid}
\end{figure}

Based on FALCO\cite{riggs2018falco1}, the ``Lightweight Space Coronagraph Simulator''
computes small linear perturbations about the nominal dark hole instead of propagating the full optical model.
It allows quickly simulating observation scenarios with time evolving \gls{WFE}, \gls{DM} drift, \gls{LOWFSC} residual jitter and \gls{HOWFSC} \cite{pogorelyuk_effects_2020}. 
The sensitivities to DM commands (the Jacobian) and to \gls{WFE} were computed in 6 wavelengths using FALCO and remain valid in the linear regime (up to 10 nm phase perturbations). 
This allows specification of DM voltages, \gls{WFE} Zernikes, LOWFS residual jitter, detector noise and switching between broadband and narrowband modes.
The Python code is designed to run fast and only requires NumPy\cite{numpy}. 
Fig. \ref{fig:leonid} shows an example dark hole time series (left) and dark hole image (right).

\subsection{CZT-based Optical Propagation}
    The Chirp Z-Transform (CZT) based optical propagation library, developed at Princeton and NASA Ames, implements 2D Fraunhofer and Fresnel diffraction propagation on arbitrarily sampled input and output planes. It provides the same functionality and answer (to within numerical precision) as the more commonly used MFT technique (Matrix Fourier Transform), but is asymptotically faster. 
    
\subsection{MSWC}

Binary stars represent a special challenge for a diffraction simulation due to the angular separation between the on-axis target and its off-axis companion. The off-axis companion can contribute stellar leakage due to optical aberrations at high-frequencies for every component in the optical train. The Multi-Star Wavefront Control (MSWC) package can be used to simulate binary stars, predict the expected leakage for different binary star imaging scenarios, and determine if the contrast leakage due to the off-axis companion introduces a background contrast floor for a particular Roman CGI imaging mode. This can flag known binaries as benign or requiring suppression. Depending on the Roman CGI imaging mode, wavefront control techniques may remove the binary companion's leakage \cite{sthomas15snwc, dsirbu2017mswc,dsirbu2018RomanMSWC}.

\subsection{WebbPSF}
WebbPSF\cite{perrin_simulating_2012,douglas_accelerated_2018-1} is a \gls{PSF} simulation tool originally developed for \gls{jwst} in Python. 
Basic \gls{SPC} modes are currently included in WebbPSF\cite{perrin_poppy_2016}\footnote{\url{https://www.stsci.edu/jwst/science-planning/proposal-planning-toolbox/psf-simulation-tool}} and are in the process of being updated to match current filters and mask designs.

\section{Yield}
\subsection{Yield Calculator}
Nemati\cite{nemati_sensitivity_2017,nemati_method_2020} developed an analytic model of instrument sensitivity that calculates the time-to-\gls{SNR} for known \gls{RV} exoplanets. 
This model has been widely used by the project team as an Excel spreadsheet and is now publicly available as part of EXOSIMS (see below).

\subsection{EXOSIMS}
EXOSIMS\cite{savransky_exosims_2017} is an open source, full exoplanet imaging mission simulation tool,  which generates a survey ensemble of possible exoplanet detections given an underlying universe (e.g. exoplanet phase curves and occurrence rates) and observatory properties such as orbit, optical system performance, and background sources.

\subsection{Imaging Mission Database}
\begin{figure}[ht]
    \centering
    \includegraphics[width=0.7\textwidth]{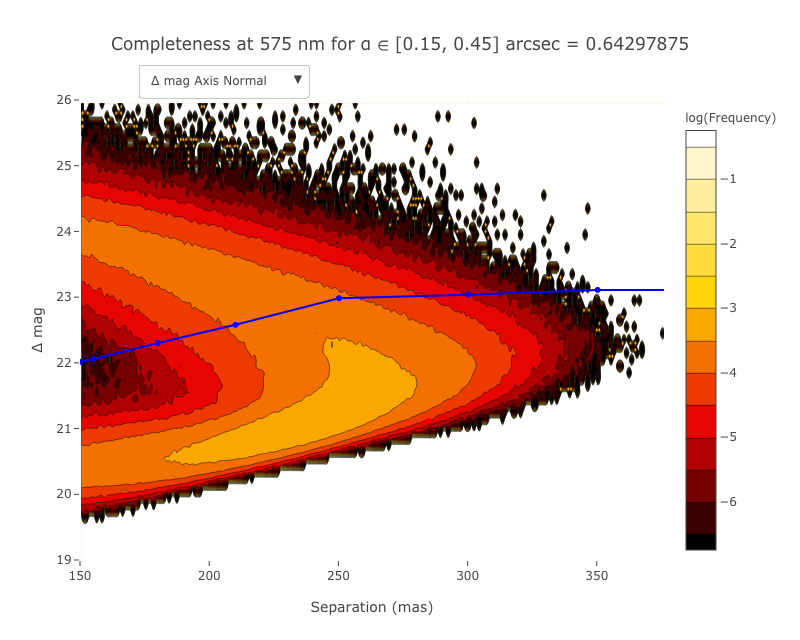}
    \caption{Map of the joint distribution of planet projected separation and $\Delta$mag for 47 Uma c generated using the Imaging Mission Database.}
    \label{fig:plandb}
\end{figure}

The online Cornell Space Imaging and Optical Systems Lab \textit{Imaging Mission Database}
uses stellar and \gls{RV} exoplanet physical properties\cite{batalha_color_2018}. As shown in Fig. \ref{fig:plandb}, these properties can be combined in joint distributions and compared to the instrumental sensitivity floor (blue curve) to assess the frequency of  detection in an observation.
Similarly, the Imaging Mission Database generates depth of search maps\cite{garrett_2017} using EXOSIMS for blind-search targets such as the  EXOCAT database\cite{turnbull_exocat_2015}.

\section{Science Targets}

\subsection{Light from Young Giant Exoplanets}
Some of the substellar companions and young giant planets discovered by ground-based direct imaging surveys will be viable targets for Roman-CGI spectroscopy and photometry. This will present the first opportunity to make observations of such objects at wavelengths shorter than 0.95 $\mu$m. At the time of writing, we have identified HD 984 B, $\beta$-Pic b, HD 206893b, HR 8799e, and 51 Eri b as possible targets for Roman-CGI bandpasses 1, 2, and 3 which include spectroscopy and photometry. HR 2562b, HR 8799d, HR 8799c, $\kappa$-And b, HD 95086b, HR 3549b, HD 1160b, and HIP 65426b  are possible targets for photometry in bandpass 4. We used \textit{coolTLUSTY} to generate self-consistent 1D radiative-convective equilibrium atmosphere models for each possible target, assuming values for effective temperature, radius, and surface gravity taken from the literature and that atmospheres are clear with solar abundances. For some targets, we also compare to models with higher metallicities and with forsterite clouds. These are discussed in detail in Lacy \& Burrows 2020\cite{lacy_prospects_2020} and are available publicly.
For all these objects, the dominant spectral features are the pressure broadened Na and K resonant doublets around 0.59 $\mu$m and 0.77 $\mu$m respectively, especially for the cooler objects where there is stronger pressure broadening at the photosphere. For the hotter objects (above $\sim$1300 K), metal hydrides like CaH and MgH have not yet condensed and rained out, so their absorption features are present in the spectra. For intermediate temperature objects and higher metallicity hot objects, absorption features from TiO and VO also appear in the spectra. When forsterite clouds are included they increase the continuum flux ratio in the optical range and weaken gaseous absorption features. Combining optical spectroscopy and photometry with NIR and MIR observations of these objects from other instruments will aid efforts to characterize these young planets since changes to cloud properties and metallicity have different effects in the optical than at longer wavelengths.  

Ideally, Roman-CGI spectroscopy will measure the strength of the $\sim$0.73 $\mu$m methane absorption feature for a reflected-light target, and constrain parameterized models incorporating cloud properties, a temperature-pressure profile, and mixing ratios of major gas-phase absorbers\cite{lupu_developing_2016,nayak_atmospheric_2017,Damiano2020}. Whether this task is achieved will depend on the nature of the planets observed, the quality of data Roman-CGI collects, and the quality of auxiliary measurements like mass and orbital separation that are available from other sources. In the event that Roman-CGI performance is insufficient to constrain detailed models, Lacy et al.\cite{lacy_characterization_2019} put forward a set of models suitable for addressing a simpler task: assessing how cool giant exoplanets compare to the cool gas giants and ice giants in our own solar system. Saturn and Jupiter have higher cloud layers and lower metallicities than Uranus and Neptune. Lower metallicities decrease the amount of methane, and higher clouds make for a smaller amount of gas above the cloud layer. Together these effects weaken Jupiter and Saturn's methane absorption features compared to Neptune and Uranus. At shorter wavelengths, from 0.4-0.6 microns, Saturn and Jupiter exhibit absorption from an unidentified chromophore which reddens their appearance. This is similar to the mix of hydrocarbons, commonly termed Tholins, that give Titan its yellow-orange appearance. Retrievals on simulated observations showed that Roman-CGI observations should be able to fit a model consisting of a two-part linear combination of Jupiter and Neptune's reflective properties. One parameter sets the short wavelength weighting towards Jupiter versus Neptune and essentially depends on whether a chromophore is present or not. A second parameter sets the longer wavelength weighting towards Jupiter versus Neptune and mainly reflects the amount of methane present above the cloud layer. This approach supposes that Jupiter and Neptune represent two bounds of cool giant planet atmosphere behavior and that those observed will fall somewhere in between. Of course, this framework is over-simplified, but, in the absence of high resolution high SNR spectra, it could provide a useful starting point. A small grid of geometric albedo models in this form is also available\footnote{\url{https://www.astro.princeton.edu/~burrows/wfirst/index.html}}, along with a GUI for calculating the light curve through out a planet's orbit and a set of models representing Jupiter's geometric albedo with constant cloud properties but varying metallicity.

\begin{figure}[ht]
    \centering
    \includegraphics[width=0.9\textwidth]{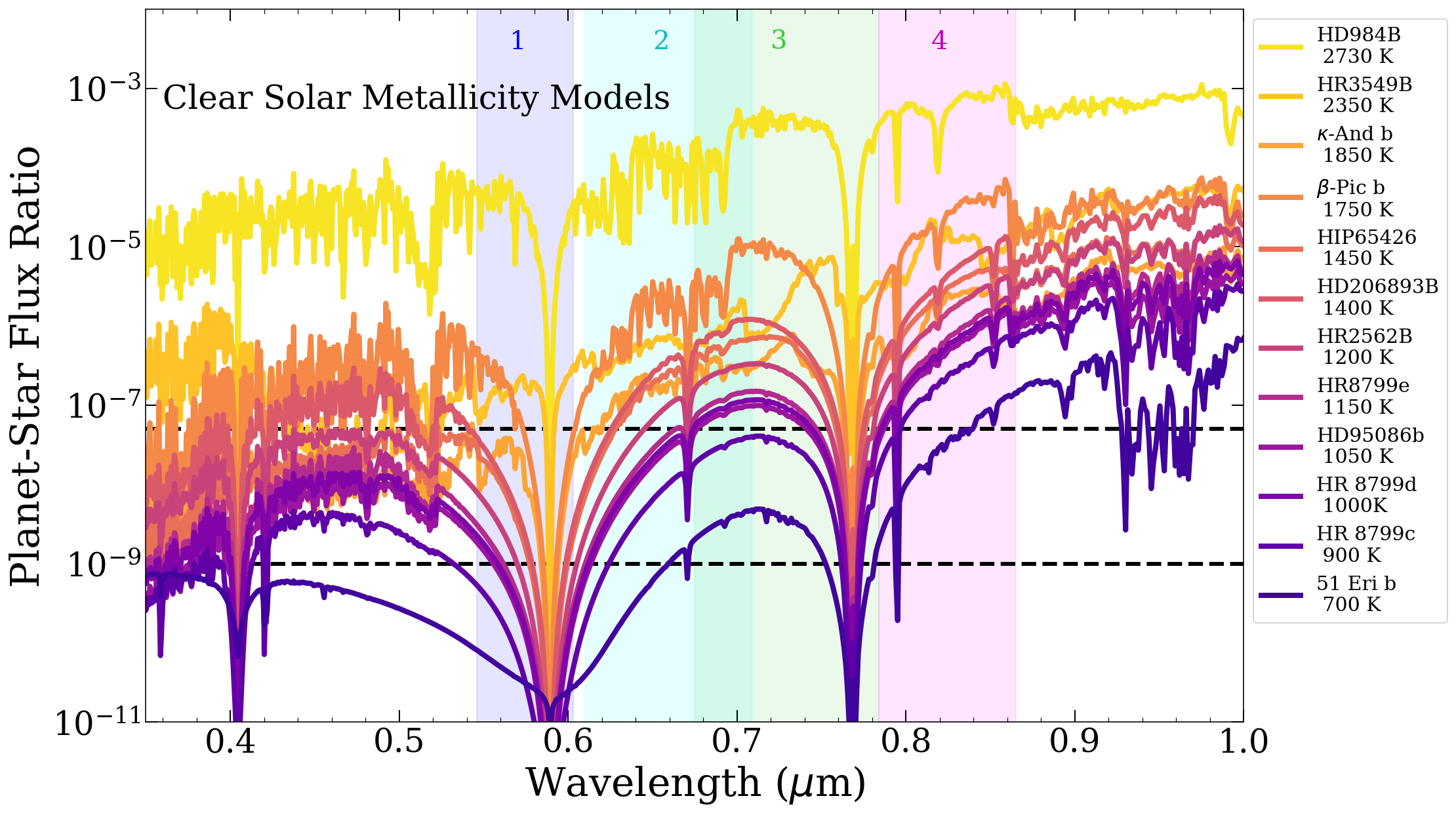}
    \caption{Model spectra for potential self-luminous targets including imaging and spectra. All models assume solar metallicity and an atmosphere free of clouds. Other model parameters are listed in table 2 of Lacy \& Burrows 2020\cite{lacy_prospects_2020}. Lines for different model spectra are colored in the order of the assumed effective temperature for each object. The shaded regions indicate the Roman-CGI bandpasses. The horizontal dashed lines mark the required contrast for technology demonstration success and the engineer's best estimate of the contrast that Roman-CGI can achieve. Note that the spectral resolution of model spectra shown here are higher than the Roman-CGI spectral resolution.}
    \label{fig:self_lum_models}
\end{figure}

\subsection{Reflected Light from Cold Gas Giants}
Giant Exoplanet Albedo Spectra and colors as a function of planet phase, separation, and metallicity from Cahoy et al\cite{cahoy_exoplanet_2010} have been released along with a reference solar spectrum.

\subsection{Debris Disks}
The  unprecedented point source sensitivity of CGI is expected to also lead to many new scattered light detections of debris disks and exozodiacal dust.
Simulations of dusty systems, particularly 1  Vir,  Eps  Eri,  HD  10647,  HD  69830,  HD 95086,  HR  8799,  and  Tau  Cet were prepared by a the Preparatory Science Project: The Circumstellar Environments of Exoplanet Host Stars by Chen et al\cite{chen_wfirst_nodate,chen_circumstellar_2014} and have been publicly released and include injected giant planets.
General libraries of disks\cite{mennesson_wfirst_2018}  with varied morphology and geometry have  convolved with coronagraph transmission functions have also been released publicly.

\subsection{Fast Extended Source Simulation with a PSF Library}

The effect of the spatially varying coronagraph transmission functions on complex scenes, such as exozodiacal dust models can be simulated using public \glspl{PSF} libraries\cite{douglas_simulating_2019}. These simulations are efficiently performed by generating an over-sampled scene model and applying a coronagraphic transfer function via matrix multiplication. Because of the field-dependent evolution of the PSF, a PSF must be generated for every angular offset of the pixel coordinates of the scene model. These PSFs are generated via interpolation of the PSFs available from IPAC and stored as matrices in memory. For example, an exozodiacal model can be flattened into a vector rather than a 2D array and multiplied by the matrix of interpolated PSFs since the coronagraph is still assumed to be a linear system. For details and examples of  disk generation, see Milani et al. \cite{milani_faster_2020}. 

\subsection{PSF subtraction}
Post-processing via \gls{KLIP} subtraction of residual speckles has been extensively explored for Roman\cite{ygouf_data_2015-1,ygouf_data_2016} and exoplanet extraction has been one of the data challenge topics\cite{mandell_wfirst_2019}.
pyKLIP\cite{wang_pyklip:_2015} supports generic data and the documentation currently includes an example of \gls{KLIP} run on an observing scenario. 





\section{Conclusions}

Roman CGI has has stimulated and nurtured a wide array of coronagraph and mission simulation tools. 
In addition to the simulation software describe above, Data Challenges\cite{hildebrandt_wfirst_2018,mandell_wfirst_2019,girard_2019_2020} have allowed the community to engage and develop additional tools\footnote{\url{https://roman.ipac.caltech.edu/sims/Exoplanet_Data_Challenges.html}}.
Integral field spectrograph modes were considered for Roman and simulated in Coronagraph and Rapid Imaging Spectrograph in Python (\verb+crispy+)\cite{rizzo_simulating_2017} which may also prove useful for other missions\footnote{\url{https://github.com/mjrfringes/crispy}}. The majority of the tools described above have been developed in Python; however, other tools have been developed for estimating coronagraph noise\cite{robinson_characterizing_2015} in languages such as IDL\footnote{\url{https://github.com/tdrobinson/coronagraph_noise}}.
It is hoped that the open-source availability of all these tools will allow the community to more rapidly develop science questions with Roman CGI and provide springboards for future coronagraphic missions.

\acknowledgments 
 This research has made use of the Imaging Mission Database, which is operated by the Space Imaging and Optical Systems Lab at Cornell University. The database includes content from the NASA Exoplanet Archive, which is operated by the California Institute of Technology, under contract with the National Aeronautics and Space Administration under the Exoplanet Exploration Program, and from the SIMBAD database, operated at CDS, Strasbourg, France\cite{wenger_simbad_2000}.

Portions of this work were supported by the Roman/WFIRST Science Investigation team prime award \#NNG16PJ24C.
Portions of this work were supported by the Arizona Board of Regents Technology Research Initiative Fund (TRIF).
\bibliography{report,wfirst} 
\bibliographystyle{spiebib} 

\end{document}